\documentstyle[aps]{revtex}
\input epsf
\epsfverbosetrue
\newcommand{\be}{\begin{equation}}
\newcommand{\ee}{\end{equation}}

\begin{document}
\draft
\title{\bf  Electron-electron  interaction in  doped GaAs at high magnetic field.}
\author{W. POIRIER,  D. MAILLY * and  M. SANQUER }
\address{CEA-DSM-DRECAM-SPEC,
C.E. Saclay,  91191 Gif sur Yvette Cedex, and * CNRS-LMM, 196 Ave. H. Ravera, 92220 Bagneux, France.}
\maketitle
\begin{abstract}
We observe an inversion of the low temperature dependence for the conductivity of doped GaAs by application of a magnetic field. This inversion happens when $ \omega_{c}\tau_{tr} \simeq 1$, as predicted by  Houghton \protect\cite{houghton} for the correction to conductivity due to screened Coulomb repulsion in the diffusive regime. This correction follows the oscillating behavior of the transport elastic time entering the Shubnikov-de Haas regime. For $ \omega_{c}\tau \ge 1$, we observe that the Hartree part of the interaction correction is suppressed. Moreover,  the total correction seems strongly reduced although its dependence stays logarithmic.  
\end{abstract}
\par Electron-electron interaction (EEI) and weak localization corrections  determine the  low temperature dependence for the conductivity of disordered metals and highly doped semiconductors.  In the two dimensional case, following Altshuler Aronov and Lee ( AAL ) \cite{altshuler}, the EEI correction to the conductivity is given in zero magnetic field and in absence of any spin relaxation by:
\be  \delta \sigma ( T  ) = { e^2 \over  2 \pi ^ {2} \hbar} ( 1 + {3 \lambda^{(j=1)} \over 4} ) ln({k_{b}T \tau\over \hbar})
\ee
where   $ \tau$ is the elastic relaxation time. The first universal term describes interaction between an electron and a hole with total spin j=0 and is  due to the exchange ( Fock ) term while $\lambda^{(j=1)}$  is  related to the direct ( Hartree ) term in the Hartree-Fock approximation of the Coulomb repulsion. In absence of any attractive virtual potential between electrons, $\lambda^{(j=1)}$ depends only on the Fermi surface and on the screening length. The exchange term  dominates the Hartree term, if the interaction potential is sufficiently smooth, i.e. its extension is larger than $\lambda_{F}$ \cite{exchange}.

\par For magnetic fields higher than $ H_{c}={ k_{b}T\over g^{*} \mu_{B}}$, the spin degenerescence is broken by Zeeman splitting, and the correction due to interaction becomes:
\be  \delta \sigma ( T  ) =  { e^2 \over  2 \pi ^ {2} \hbar} ( 1 + { \lambda^{(j=1)}\over 4} ) ln({k_{b}T \tau\over \hbar})
\ee
\par

 The expressions (1) and (2) are valid for a diffusive motion \cite{rudin}  and are modified  when the cyclotron frequency $ \omega_{c} = { m^{*} \over e H }$ is comparable to the elastic relaxation time $ \tau$ .
\par In this classically high magnetic field case, it is known that the tensor of  conductivities is anisotropic:
\be\cases{\sigma_{xx}  = {1 \over1 + (\omega_{c} \tau)^{2}}\sigma(\omega_{c}=0) \cr \sigma_{xy}  = {- \omega_{c} \tau \over 1 + (\omega_{c} \tau)^{2}} \sigma( \omega_{c}=0 )\cr }
\ee
\par Houghton \cite{houghton} has shown that: 
\be\cases{\delta \sigma_{xx} = \delta \sigma( \omega_{c}=0 ) \cr
\delta \sigma_{xy} = 0} \ee
The equation (4) is a general result which is valid for any dimensionality and any kind of interaction between electrons \cite{altshuler}.
A a consequence the  correction to the conductivity is :
$\delta \sigma( \omega_{c}  ) = \delta ({ \sigma_{xx}^{2} + \sigma_{xy}^{2} \over \sigma_{xx} }) = \delta \sigma( \omega_{c} = 0 ) ( 1 - (\omega_{c} \tau)^{2})$, despite
$\sigma( \omega_{c}  ) = \sigma( \omega_{c} = 0 ) $.
 With equation (2), one finally obtains :
\be \delta G = \delta \sigma =   { e^2 \over  2 \pi ^ {2} \hbar} ( 1 + {\lambda^{(j=1)}  \over 4} ) ( 1 - (\omega_{c} \tau)^{2}) ln({k_{b}T \tau \over \hbar})\ee
This remarkable result is valid for $\omega_{c} \tau \le\ge 1$ as demonstrated in ref. \cite{houghton}\cite{houghton2}. It means that the logarithmic correction to the conductance due to interaction increases steadily as a function of magnetic field, changing its sign at $\omega_{c} \tau = 1$. Equation 5 shows also that  oscillations of  $\tau(H)$ with magnetic field (in the Shubnikov-De Haas regime) may give oscillations of $\delta G$ .
\par The aim of this work is twofold: first we will show a direct experimental observation of the inversion of the correction. We will confirm the temperature (ln(T)) and magnetic field $(1-(\omega_{c}\tau)^{2})$ dependences according to equation 5. Then we will use this fact to extract unambiguously the $(1+{\lambda^{(j=1)}\over 4})$ term in  both low and high classical magnetic fields, and find that the Hartree contribution is suppressed (~$\lambda^{(j=1)} \simeq 0$~) once  $\omega_{c} \tau \ge 1$. To our knowledge such a result have been never reported up to now. Moreover, the substraction of the  Shubnikov-de Haas oscillations permits us to show that the logarithmic term in equation 4 is effectively given by $ln({k_{b}T \tau_{tr}\over \hbar})$ at low magnetic field
where $\tau_{tr}$ is the transport relaxation time. However,the amplitude of this factor seems strongly  reduced at high field.
\par   Equation 5 means that  $ \delta \sigma $ increases in amplitude as the square of the magnetic field, leading eventually to an Hall insulating state characterized by $\sigma_{xx} \simeq 0$ and $\sigma_{xy} $ constant. This prediction has been studied by Murzin et al. \cite{murzin} in 3D doped semiconductors at high magnetic field. But the crossover at $ \omega_{c} \tau \simeq 1 $ has not been studied.  A merit of our samples is to conjugate a relatively high disorder, giving a large EEI correction to the conductivity even at small fields, and a classical high magnetic field regime above 3 teslas. Electron interactions have been also studied in 2D high mobility GaAs heterostructures by Choi et al. \cite{choi}. They observed that the correction to conductivity due to interaction varies like $(\omega_{c} \tau)^{2}$ ( for $\omega_{c} \tau \ge 1$ and $ T \ge 1K$), and use that fact to study extensively the amplitude of the correction for various geometries. Our experiment differs from ref. \cite{choi} because the samples are in the diffusive regime, where AAL theory is applicable. In addition the Shubnikov-deHaas oscillations do not depend on temperature in our sample, because the elastic mean free path is much smaller  ( large Dingle temperature ) than in ref. \cite{choi} ( and the experiment is performed at lower temperature ). This permits us to extract the temperature dependence of the correction and not only the associated magnetoresistance. For this limitation the sign inversion predicted in equation 4 is not seen in ref \cite{choi}.

\par We have used MBE grown  GaAs doped at $ 2.2 \ \ 10^{23} Si \ \ m^{-3}$.  Because our samples are based on a 300nm thick layer, in the low temperature regime considered,  samples are effectively two dimensional: both the phase breaking length and the thermal length $ L_{T} = \sqrt{ h \over e k_{B}T }$ are larger than the thickness below 1 kelvin. A $250 \times 200 {\mu m}^2$ sample with ohmic AuGeNi contacts is defined by etching. The system is characterized by the following parameters:
$D=3.2 \ \ 10^{-3} m^{2}s^{-1}, k_{f}l=6.5, \tau_{tr}=1.01 \ \ 10^{-13} s, E_{f}=240 K, 
a_{b}=95 \AA $ and $R_{c} = 492 $ Ohms is the resistance per square.
\par
\begin{figure}[h]
\centerline{\epsfxsize=12cm \epsfbox{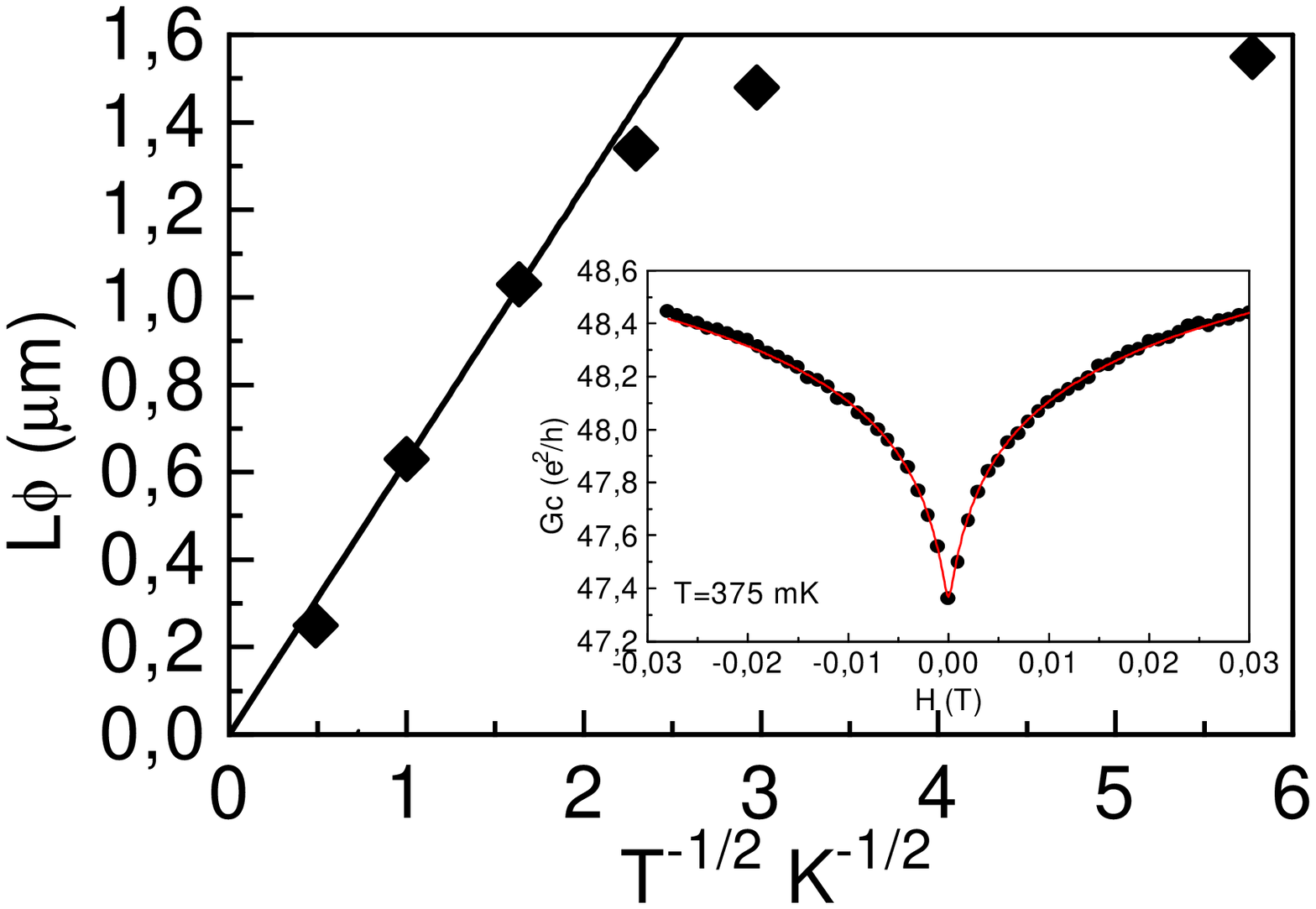}}
\vskip 1cm
\caption{ $L_{\phi}$ versus temperature. The solid line is the prediction by equation 7.  Inset: the low field magnetoconductance at T = 375 mK with the weak localization fit.}
\label{fig:lphi}
\end{figure}

 To separate the EEI correction, we first analyse the weak field magnetoconductance which is entirely due to the weak localization correction (see  inset of figure \ref{fig:lphi} )\cite{altshuler}:
\be 
\delta\sigma(H) = {{e^2} \over {2{\pi^2} \hbar}} f_2 (2{({{{L_\phi} \over L_H}})^2})
\ee
\par with $f_2(x) = ln x + \psi(x+{1\over2})$. $L_{H} = \sqrt{ \hbar / e H}$  is the magnetic length and $\psi(x)$ the digamma function.

 Figure \ref{fig:lphi} shows the temperature dependence of the phase breaking length  $L_\phi = \sqrt{D\tau_{\phi}}$ (~D is the diffusion constant and $\tau_{\phi}$ the phase breaking time~).
For electron-electron interaction in two dimensions, Altshuler et al. \cite{altshuler} obtain:
\be  L_\phi = \sqrt {2\pi D {{\hbar}^2} \over {{k_b} {e^2} R_{c} ln ({{\pi \hbar} \over {e^{2} R_{c}}})}} \ \ { {T^{-{1 \over2}}}} \ee
\par With the measured sample parameters, we find $L_\phi (\mu m) = 0.63 \ \ T^{-{1 \over2}} $ in excellent agreement with the weak localization measurement between  150 mK and 4K ( see figure  \ref{fig:lphi}). At very low temperature (~$T \le 150 mK$~) a saturation is nevertheless observed, attributed either to high frequency heating, to dephasing due to magnetic impurities or to general electromagnetic environment considerations \cite{webb}.
\par  In the intermediate magnetic field regime (~$0.02 \le H \le 0.5$~) both Zeeman effect and weak localization  give non negligible and opposite contributions to the magnetoconductance. In addition a crossover in the effective dimensionality occurs when the magnetic length is comparable to the sample thickness. For these reasons we do not fit the magnetoconductance in this intermediate regime. From the temperature dependence of the conductance correction both in zero magnetic field and for $H=1T$, we determine selfconsistently $\lambda^{(j=1)}$ (~see inset of figure 2~). In fact, above $H=1T$ the weak localization contribution is negligible and the Zeeman level degeneracy breaking is effective for our lowest electron temperature. The conductance correction should obey to eq. 5, i.e. the slope of $\delta G({e^2 \over h})$ versus $\delta lnT$ (~divided by $ 1 -{(\omega_{c}\tau(H))}^2$, that is 0.929 at H=1T~) is given by ${1 \over \pi}{1 \over 1.3} ( 1 + { \lambda^{(j=1)}\over 4} )$ (~where the factor d=1.3 corresponds to the the length divided by the width of the sample~). At zero magnetic field the same slope is given by ${1 \over \pi}{1 \over 1.3} ( 1 + { 3 \lambda^{(j=1)}\over 4} + 1 )$, where the last factor 1 is due to the weak localization term  ${1 \over \pi}{1 \over 1.3} {e^{2} \over h} ln(\tau_{\phi}/ \tau)$ (~with  $\tau_\phi \propto T^{-1}$~). The first evaluation gives  $\lambda^{(j=1)} \simeq -1.55 +- 0.1$ while the second estimation is compatible with  $\lambda^{(j=1)} \simeq -1.2$. This corresponds to a strong screening case in d=2. The small discrepency may be related to a small spin splitting a zero magnetic field \cite{sos} or to additionnal terms, for instance the Maki-Thomson term.
. \par Moreover, the absolute magnetoconductance between $H=0T$ and $H=1T$, is well accounted by balancing the the weak localization suppression and the Zeeman splitting effects (~see inset of figure 2~):
\be G(1T)-G(0) \simeq {1 \over \pi}{1 \over 1.3} {e^{2} \over h}( ln(\tau_{\phi}/ \tau_{tr})  + {\lambda^{(j=1)} \over 4} ln({k_{B}T\tau_{tr}\hbar})  ) \ee
For instance at $T = 1K$, we find $G(1T)-G(0) \simeq  1.16 {e^{2} \over h}$ and we estimate $G(1T)-G(0) \simeq  1.34 {e^{2} \over h}$ (T=1K, $\tau_{\phi} = 1.32 \ \ 10^{-10}s$ and $\lambda^{(j=1)}=-1.55$~).

Our value  $\lambda^{(j=1)}$ corresponds to a screening larger than the estimation based on the Thomas-Fermi approximation \cite{altshuler}, but is not surprizing considering the relatively high carriers concentration.
\par After elimination of the weak localization and Zeeman splitting effect, it is possible to investigate precisely the correction due to interaction above $H = 1T$. First one has to consider the $ ( 1 - (\omega_{c} \tau)^{2}) $ term in equation 4, which give two main effects: a change of sign for $\delta G(T)$ as $\omega_{c} \tau \simeq 1$ and oscillations of $\delta G(T)$ resulting from oscillations of $\tau(H)$.

\begin{figure}[h]
\centerline{\epsfxsize=12cm \epsfbox{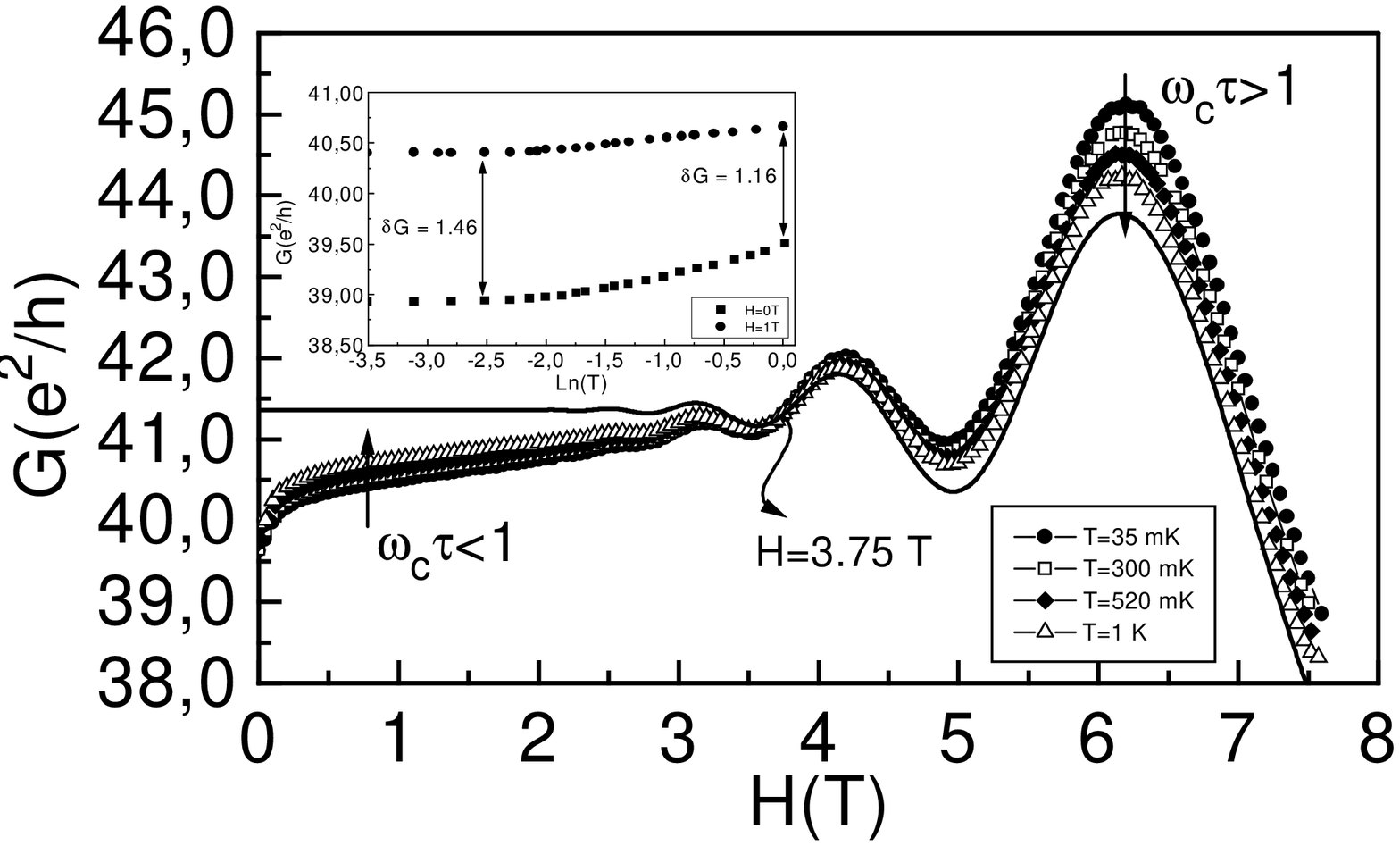}}
\vskip 1cm
\caption{  G(H) at various temperatures. The solid line is the fit of the Shubnikov-deHaas oscillations. Inset: the correction of $G$ versus T at two magnetic fields (~squares: H=0T ; circles: H=1T~). The absolute variation between these two fields is given by the weak localization and the Zeeman splitting terms.}
\label{fig:gth}
\end{figure}

\begin{figure}
\centerline{\epsfxsize=12cm \epsfbox{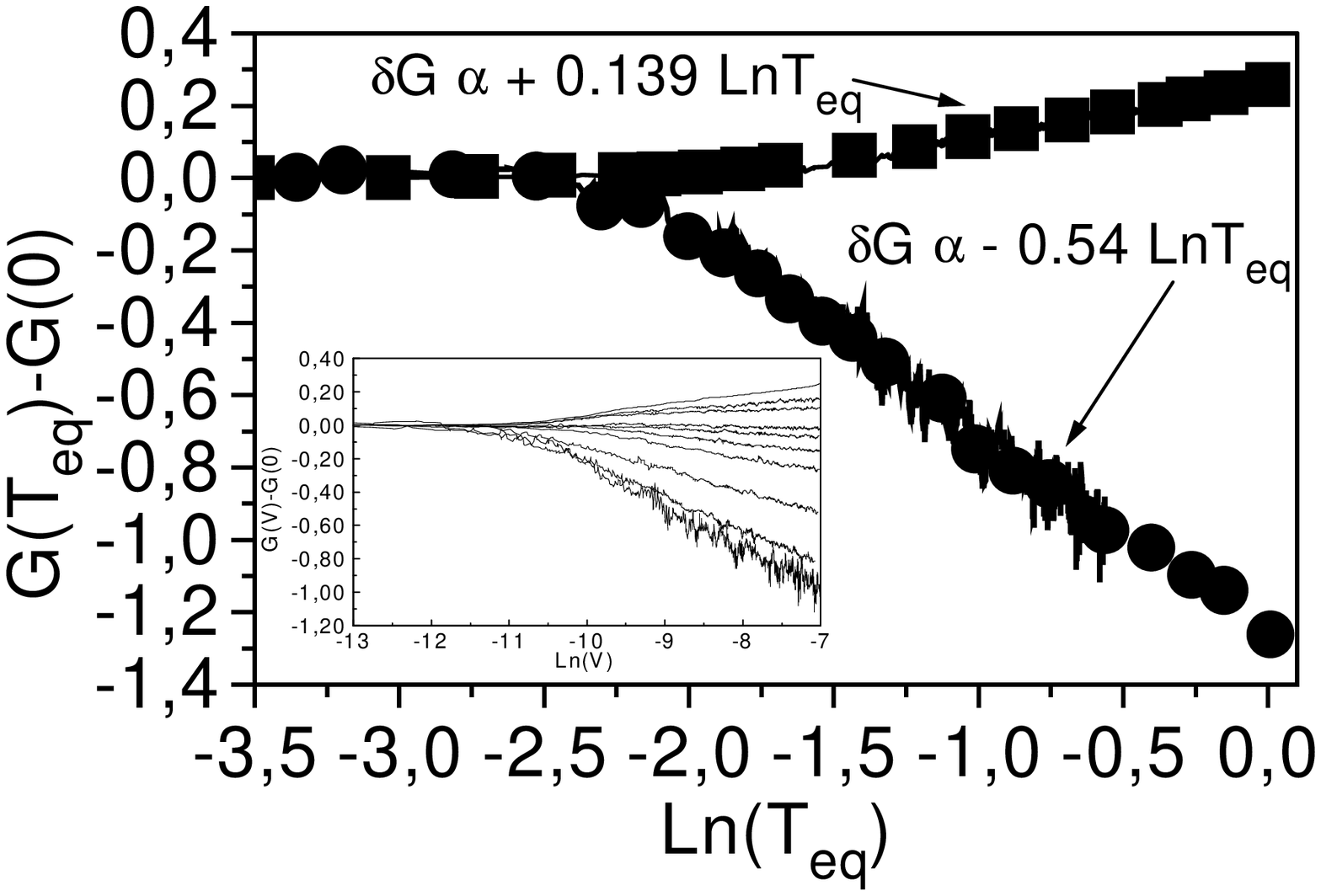}}
\vskip 1cm
\caption{Conductance versus temperature for H=0.5T (~squares~) and H=6.6T (~circles~). $T_{eq}$ is the measured temperature for measurements in the linear regime or the temperature deduced from the bias-temperature relation discussed in the text for non linear conductance measurements (~solid lines~). Inset: G(V) at various intermediate magnetic fields: 1,1.5,3,3.75,4,4.5,5,5.5,6,6.6T. (~top to bottom~)}
\label{fig:log}
\end{figure}

Figure  \ref{fig:gth} shows the absolute magnetoconductance at various temperatures. One can see the change in the temperature dependence of the conductance at a magnetic field of about 3.75 teslas. This is confirmed in the Figure \ref{fig:log} which details  the correction to the conductivity versus temperature for two  magnetic fields: H=1T and H=6.6T. The correction varies like the logarithm of the temperature as predicted by equations 1 and  5.
Note that the cancellation of the correction at H=3.75T permits to determine precisely the Drude conductance: $G_{Drude}=41.36 ({e^{2}\over h})$.
 \par The figure \ref{fig:log} shows also the conductance versus  bias at $T \simeq 100 mK$ after rescaling the voltage as an effective temperature $ T'=  \beta V^{2 \over 5}$ ($\beta(H=1T)=15$, $\beta(H=6.6T)=10$; this change is not explained~).
$T'$ is much lower than ${eV \over k_{b}}$ because the sample dimensions are much larger than the electron-phonon coupling length.
$T'$ is approximately given by
\be T' \simeq {({T_{0}^{5}}+ {{V^2}\over{\sum \varrho {L^2}}})}^{1 \over 5} \propto V^{2 \over 5}
\ee
with $\varrho$ the resistivity and $\sum$ is estimated to be $ 3.9 \ \ 10^{-4} nW {\mu m^{-3}} {K^{-5}}$. This value of $\sum$ could be compared with its theoretical expression \cite{urbina}: $\sum = 0.524  \alpha \gamma$ with
${\tau_{ep}}^{-1}={\alpha T'^{3}}$ and $\gamma={{{\pi^2}\nu {k_{B}^2}} \over3}$.
$\nu$ is the density of states and  $\alpha$ is a numerical model dependent constant.
\par
\begin{figure}[h]
\centerline{\epsfxsize=12cm \epsfbox{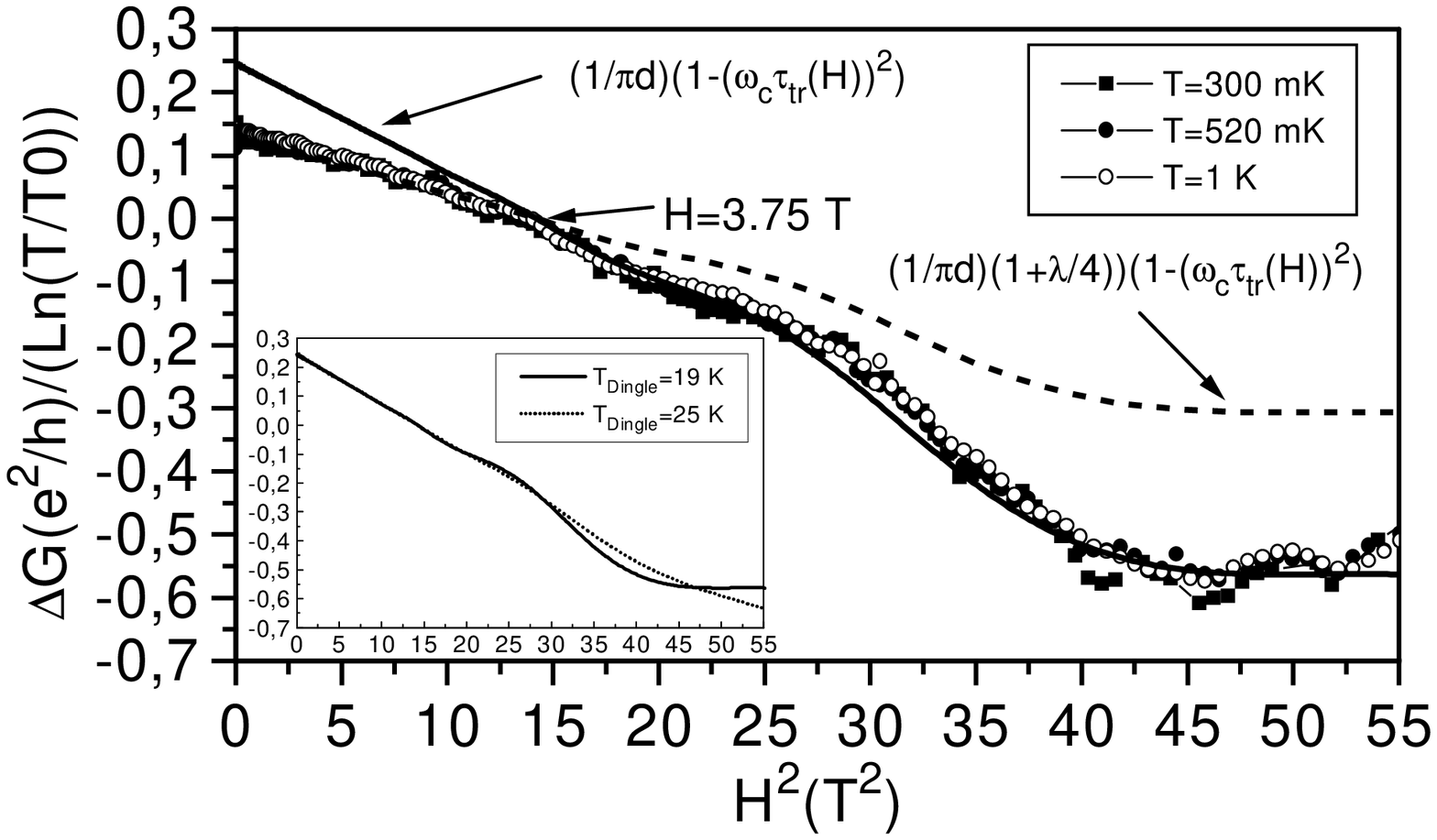}}
\caption{The correction to conductivity due to electron interaction versus $H^2$. The base temperature $T_{0}$ is 150 mK; the symbols refer to different temperatures. The dashed and solid lines are respectively the equation 5 with $\lambda^{j=1} \simeq -1.55$ (~excellent at low field~) and with $\lambda^{j=1} = 0$ (~excellent at high field~). Inset: Two fits with  $\lambda^{j=1} = 0$ for two different Dingle temperatures showing the sensitivity to this parameter. } 
\label{fig:factor}
\end{figure}

 We observe that the change in the temperature or bias dependence of the conductivity happens precisely when $ \omega_{c}\simeq \tau_{tr}^{-1}$ where $\tau_{tr}$ is the transport relaxation time (~see figures  \ref{fig:gth} and \ref{fig:factor}~). In that range of fields, the sample exhibits pronounced Shubnikov-de Haas oscillations periodic in $ 1 \over H $ ( see  figure \ref{fig:gth} ), which permits us to determine  the thermodynamic relaxation time $\tau_{thermo}$ to be $6.410^{-14} s$ \cite{sdh}. We find that  $\tau_{tr} \simeq 1.5 \tau_{thermo}$. The diffusion by impurities is quite isotropic. Moreover, this time corresponds to a large value of the Dingle temperature: $T_{D}\simeq 19 K$.
This  value much larger than our experimental range [0.1 K , 1 K ] implies that the temperature changes of the conductance  are strickly related to the EEI effects.
This permits us to investigate the absolute values for exchange and direct terms (~see figure \ref{fig:factor}~). Indeed, by substracting conductance versus magnetic field at different temperatures T and T' we can estimate the term: $\delta G  = G(T)-G(T') = ({1\over{\pi 1.3}}) ( 1 + {\lambda^{(j=1)}  \over 4} ) ( 1 - (\omega_{c} \tau)^{2}) ln(T/T')$ . From the low field analysis we have obtained $\lambda^{(j=1)} \simeq -1.55 +- 0.1$, that corresponds to a relatively strong screening case, which makes the direct term comparable to the exchange term.  Figure \ref{fig:factor} shows that this estimation is valid up to $\omega_{c} \tau_{tr} \simeq 1$. But, as $\omega_{c} \tau_{tr} \gg 1$, the fit deviates strongly from the data. In this high magnetic field regime, we find that 
\be 1.3 \delta G = \delta \sigma =   { e^2 \over  2 \pi ^ {2} \hbar} ( 1 - (\omega_{c} \tau_{tr})^{2}) 
 ln(T/T'))\ee
without any adjustable parameter: the direct term is destroyed (~$\lambda^{(j=1)} \simeq 0$~) and the correction is just given by the exchange part, qualitatively as if the screening becomes much less efficient.
\par The reason for the cancelation of the Hartree term needs to be clarified, taking into account that it happens as $\omega_{c} \tau \ge 1$. That suggests an orbital effect, perhaps due to the reinforcement of the forward scattering as compared to the backward ones: as $\omega_{c} \tau \gg 1$, the backward scattering $\Delta k \simeq 2k_{F}$
is diminished as compared to the forward scattering $\Delta k \simeq 0$. The direct ( resp. exchange) correction is proportionnal to $\Delta k \simeq 2k_{F}$
(~resp. $\Delta k \simeq 0$~), that could explain our new experimental observation.

\par To complete our analysis, we have  substracted the Shubnikov-de Haas fit in order to extract the total correction to the conductance i.e. to evaluate the absolute value of the $ln({k_{b}T \tau \over \hbar})$ term in equation 5. This is possible because of the excellent evaluation obtained for the other terms. At weak field, we verify that  the absolute value of the correction  agrees perfectly with the prediction of the Equation 5 with $\tau = \tau_{tr}$.  But at higher magnetic field (~$\omega_{c} \tau_{tr} \ge 1$~), the equation 5 predicts a larger correction than the one measured. A quantitative agreement is obtained if the term (${k_{b}T\tau_{tr} \over\hbar}$) is multiplied by a factor 50. Note that this factor does not enter in the relative $\delta G  = G(T)-G(T')$ measurement. This result suggests that departure from the diffusive regime-strictly valid only at low magnetic field- is accompagnied by a strong absolute reduction in amplitude for the correction due to electron-electron interaction.

 \par In conclusion, our diffusive GaAs sample exhibits large corrections due to disorder and interaction in zero magnetic field. Above $H \ge 1T$ only interaction corrections due to exchange  and Hartree terms of the screened Coulomb repulsion persist. These corrections leads to ${\delta \sigma \over \delta T } < 0$ at low temperature. When a high magnetic field is applied such that  $\omega_{c}\tau_{tr} \simeq 1$, the temperature dependence  changes its sign leading to ${\delta \sigma \over \delta T } > 0$, as predicted by Houghton at al. \cite{houghton}. The whole functional dependence of the correction  in $ 1 - {\omega_{c}\tau_{tr}(H)}^2$ is obtained, including the  the Shubnikov-deHaas oscillations of $\tau(H)$. We have been able to normalize the magnetoconductance curves at various temperatures and we show that the Hartree term is canceled when $\omega_{c}\tau_{tr}(H) \ge 1$. Moreover we have measured the absolute value for the interaction correction. Its predicted dependence is verified at low magnetic field, but when $\omega_{c}\tau_{tr} > 1 $ it is strongly reduced. 

\par We aknowledge B. Etienne for providing the MBE GaAs:Si layers, and  R. Tourbot for its technical support and V. Falko for fruitfull discussions.


\begin{references}

\bibitem{houghton} A. Houghton, J.R. Senna and S.C. Ying, Phys. Rev. B25, 2196 ( 1982 ).

\bibitem{altshuler} B. L. Altshuler et al Physics Reviews Vol 9, p225 , Soviet Scientific Reviews Section A, edited by I.M. Khalatnikov ( 1987 );
B. L. Altshuler and  A.G. Aronov " EEI in disordered conductors ". B. L. Altshuler and A.G. Aronov, Sov. Phys. JETP 50 (5), 968 (1979 ).


\bibitem{exchange}  In that case the Fourier component of the potential at zero wave vector is much larger than the mean Fourier component at twice the Fermi wave vector:
$ V(0) \gg V(2k_{F}) $
The sign of interaction correction is given by the sign of $ V(0)-2V(2k_{F})$, where the factor 2 is due to the fact that the exchange interaction, contrarily to the direct interaction, is between electrons with the same spin.

\bibitem{rudin} Rudin et al. ( A.M. Rudin, I.L. Aleiner and L.I. Glazman, cond-mat 9603122 )
 has recently studied theoretically the nearly ballistic regime for equation 1.

\bibitem{houghton2} A. Houghton, J.R. Senna and S.C. Ying, Phys. Rev. B25, 6468 ( 1982 ).


\bibitem{murzin} S. S. Murzin and A. G. M. Jansen, J. Phys.: Condens. Matter 4, 2201 ( 1992).


\bibitem{choi} K. K. Choi, D. C. Tsui and S. C. Palmateer, Phys. Rev. B33, 8216 ( 1986 ).

\bibitem{webb} Mohanty et al. Phys. Rev. Lett. 78, 3366 ( 1997 ).


\bibitem{sos} A strong band spin-orbit coupling exists in GaAs, which could imply a (partial) spin degeneracy even in zero magnetic field. This complicates the observation in GaAs of the  magnetoconductance associated to the Zeeman effect in the diffusion channel. To our knowledge this magnetoconductance effect has not been reported up to now (~see for instance \protect\cite{choi}, where Zeeman splitting is supposed to happen at  magnetic fields higher than studied~).


\bibitem{urbina} F. C. Wellstood, C. Urbina and J. Clarke, Appl. Phys. Lett. 54, 2599 ( 1989 ). 

\bibitem{sdh} p1031 in "Landau level spectroscopy" vol. 27-2, Landwehr-Rashba eds. Agranovitch-Maradudin Gen. eds. North Island, 1991.

\end{references}
\end{document}